\documentclass[12pt]{article}

\usepackage{amsmath,amssymb}
\usepackage{latexsym}
\usepackage{bm}
\usepackage{graphicx}

\newcommand{\R}{{\mathbb R}}

\newcommand{\bee}{\begin{equation*}}
\newcommand{\eee}{\end{equation*}}
\newcommand{\be}{\begin{equation}}
\newcommand{\ee}{\end{equation}}

\begin{document}

\markboth{A. G. Ramm}{      }

\title{
Does negative refraction make a perfect lens?}

\author{A. G. Ramm$\dag$\\
\\
$\dag$Mathematics Department, Kansas State University,\\
Manhattan, KS 66506-2602, USA\\
email: ramm@math.ksu.edu\\
}

\date{}
\maketitle

\begin{abstract} \noindent 
A discussion of a question, studied earlier by 
V.Veselago in 1967 and by J. Pendry 
in 2000, is given. The question is:  can a slab of
the material with negative refraction make a perfect lens? 
Pendry's conclusion was: yes, it can. Our conclusion is: no, in
practice it cannot, because of the fluctuations of the refraction
coefficient of the slab.
Resolution ability of linear isoplanatic optical instruments
is discussed. 
\\
{\bf key words}:   Negative refraction, EM waves, wave propagation,
resolution ability
\end{abstract}

\section{Introduction\label{s1}}
Negative refraction has been studied extensively (see \cite{A} and
references therein). In \cite{P} the following question is
discussed. Assume that a plane electromagnetic (EM) wave
$E=e_2e^{i(k_1x+k_3z)}:=e_2e^{ik_1x}u_0(z)$, $u_0(z)=e^{ik_3z}$, is 
incident from the region $z<0$ onto
an infinite slab (layer) $0<z<d$, filled with a material with negative
refraction (negative index). This means that $\epsilon<0$ and $\mu<0$
in the slab, where $\epsilon$ and $\mu$ are constant dielectric and
magnetic parameters. The refractive index 
$$n=(\epsilon
\mu)^{1/2}=(|\epsilon||\mu|e^{2i\pi})^{1/2}=-(|\epsilon||\mu|)^{1/2}<0.$$
The time dependence is given by the factor $e^{-i\omega t}$ and this 
factor is omitted throughout. By $\omega>0$ the frequency is denoted.
The vectors $\{e_j\}_{j=1}^3$ form a Cartesian orthonormal basis of
$\R^3$. Outside the slab, in the region $z<0$ and $z>d$, we assume that
$\epsilon=\epsilon_0$ and $\mu=\mu_0$, where $\epsilon_0=\mu_0=1$.
 It is assumed in \cite{P} that in 
the slab $\epsilon=-1$ and $\mu=-1$, so that 
$n=-1$ in the region $0\leq z\leq d$. By $k_j$ the components of the wave 
vector
$\vec{k}$ are denoted,  $\vec{k}=k_1e_1+k_3e_3$,
$|\vec{k}|^2=|k_1|^2+|k_3|^2=\omega^2\epsilon\mu.$ In our case $k_2=0$,
and $\epsilon\mu>0$. Let us 
denote the constants
$\epsilon$ and $\mu$ in the slab by $\epsilon_1$ and $\mu_1$. 

The
governing equations are the Maxwell's equations: 
\be\label{eq1} \nabla \times E=i\omega \mu
H,\quad  \nabla \times H=-i\omega \epsilon E, 
\ee 
and 
the tangential components of the electric and magnetic fields $E_t$ and 
$H_t$ are continuous across the interfaces $z=0$ and $z=d$. 


Our calculations prove
the following conclusion: 

{\it If $n=-1$ in the slab, then the image,
of a point source, located at the point $(0,0,-f)$, is a
point $(0,0,2d-f)$.} 

This conclusion agrees with the one obtained earlier in
\cite{V} and \cite{P}. In \cite{P} only 
the propagation of plane waves is discussed. However, the point source is 
a linear
combination of the plane waves (including the evanescent waves), 
so there is no loss of generality in considering propagation of the 
plane waves. For
example, a well known representation of a point source
as a linear combination of the plane waves is the formula:
$$\frac{e^{ik|x|}}{4\pi|x|}=\frac{1}{2\pi
i}\iint_{-\infty}^\infty
\frac{e^{i(k_1x_1+k_2x_2+x_3\sqrt{k^2-k_1^2-k^2_2})}}{\sqrt{k^2-k_1^2-k_2^2}}dk_1dk_2,$$
where $x_1=x,$ $x_2=y$, $x_3=z$, and $\sqrt{k^2-k_1^2-k_2^2}>0$ if
$k_1^2+k_2^2<k^2$, $\sqrt{k^2-k_1^2-k_2^2}=i\sqrt{|k^2-k_1^2-k_2^2|}$
if $k_1^2+k_2^2>k^2.$ 

Let us derive the formulas for the electromagnetic
field, propagating through the slab,
 and prove the  conclusion, formulated above.

Let us look for a solution to \eqref{eq1} 
of
the form: \be\label{eq2} E=e_2e^{ik_1x}u(z),\ u(z)=\left\{
                                                     \begin{array}{ll}
                                                       
e^{ik_3z}+re^{-ik_3z}, & \hbox{$z<0$,} \\
                                                       
Ae^{ik_3'z}+Be^{-ik'_3z}, & \hbox{$0\leq z\leq d$,} \\
                                                       te^{ik_3z}, & \hbox{$z>d$,}
                                                     \end{array}
                                                   \right.
 \ee where the four coefficients $r, A,B,t,$ are to be found from the
four boundary conditions: 
\be\label{eq3} 1+r=A+B,\quad
k_3(1-r)=\frac{k_3'}{\mu_1}(A-B), \ee \be\label{eq4}
Ap'+\frac{B}{p'}=tp,\quad
\frac{k_3'}{\mu_1}\left(Ap'-\frac{B}{p'}\right)=tk_3p .
\ee 
Here
\be\label{eq5} p':=e^{ik'_3d},\quad p:=e^{ik_3d},\ee

\be\label{eq6}
k_3':=\left\{
          \begin{array}{ll}
            \sqrt{\omega^2\epsilon_1\mu_1-k_1^2}>0, & \hbox{$k_1^2<\omega^2\epsilon_1\mu_1,$} \\
            i\sqrt{|\omega^2\epsilon_1\mu_1-k_1^2|}, & 
\hbox{$k_1^2>\omega^2\epsilon_1\mu_1,$}
          \end{array}
        \right.
\ee and the boundary conditions \eqref{eq3} and \eqref{eq4} come
from the continuity of the tangential components of the electric 
and magnetic fields, that is, $[e_2,E]$ and $[e_3,H]=\frac{[e_3,\nabla
\times E]}{i\omega \mu}$, across the planes $z=0$ and $z=d$,
respectively. Here $[e,E]$ denotes the vector product of two vectors. The
four equations \eqref{eq3}, \eqref{eq4} for the four unknowns
$r,A,B,t,$ can be solved analytically: 
\be\label{eq7}
r=\gamma\frac{1-p'^2}{\gamma^2p'^2-1},\quad
t=\frac{p'}{p}\cdot \frac{\gamma^2-1}{\gamma^2p'^2-1}, \ee \be\label{eq8}
A=\frac{1-b}{1+b}\cdot\frac{\gamma^2-1}{\gamma^2p'^2-1},\quad
B=\frac{1-b}{2}\cdot\frac{\gamma^2-1}{\gamma^2p'^2-1}, \ee where $p'$ and
$p$ are defined in \eqref{eq5}, and 
\be\label{eq9}
\gamma:=\frac{1-b}{1+b},\quad b:=\frac{\mu_1k_3}{\mu_0k_3'}. \ee If
$\epsilon_1=\mu_1=-1$, then $k_3'=k_3$, $b=-1$, $\gamma=\infty$,
\be\label{eq10} r=0,\ t=e^{-2ik_3d},\ A=0,\ B=1.\ee 
The coefficients
$r$ and $t$ are called the reflection and transmission coefficients,
respectively. In formulas (12), (21) in \cite{P} by
$T$ and $T_s$ not the standard transmission coefficients are
denoted, but the values $u(z)$ at $z=d$. The values of the field
everywhere in the space are given by formulas \eqref{eq2},
\eqref{eq5}-\eqref{eq9}.

\section{A discussion of the exact solution}
Let us draw some physical conclusions from the the basic formulas
\eqref{eq2}, \eqref{eq5}-\eqref{eq9}. First, consider the simplest
case $\epsilon_1=\mu_1=-1$. Then formulas \eqref{eq2}, \eqref{eq10}
hold. These formulas imply that a ray, passing through a point
$(0,0,-f)$, $f>0$, with parameters $k_1>0$, $k_3>0$, $k_2=0$, will have
parameters $(k_1,0,-k_3)$ after passing through the interface $z=0$,
and again the parameters $(k_1,0,k_3)$ after passing through the
interface $z=d$. In this geometrical argument it is assumed that
$d>f$. The ray is the straight line orthogonal to the wave front of
the plane wave. The ray issued from the point $(0,0,-f)$ with
parameters $(k_1,0,k_3)$, hits the plane $z=0$ at the point
$(h_0,0,0)$, where $h_0=f\tan \theta_0,\
\tan \theta_0=\frac{k_1}{k_3},$ and $\theta_0$ is the acute angle that
the ray makes with $z$-axis. The $z$-axis is perpendicular to the
slab. After passing the plane $z=0$, the ray hits the $z$-axis at the
point $(0,0,f)$ and then it hits the line $z=d$, $y=0$, at the point
$(-h_1,0,0),$ where $h_1=(d-f)\tan\theta_0$. Finally, after passing
the plane $z=d$, the ray hits the $z$-axis at the point $(0,0,2d-f)$.

The conclusion is :
 
{\it Regardless of the size of $\theta_0$, all the
rays, passing through the point $(0,0,-f)$, intersect the $z$-axis at
the point $2d-f$, $d>f>0$.}

 This means that the slab with
$\epsilon_1=\mu_1=-1$ acts like a "perfect lens", as is claimed in
\cite{V} and \cite{P}. There is no need to discuss separately the case 
of evanescent waves, i.e., the case when
$\omega^2\epsilon_1\mu_1<k_1^2$. Our derivations are different from
those in \cite{P}, but the conclusion in the case
$\epsilon_1=\mu_1=-1$ is the same.

However, in
\cite{P} he following questions are not discussed: 

{\it 1) One cannot
have $\epsilon=-1$ and $\mu=-1$ exactly due to defects in the
material of the slab. Will the above conclusion, (namely, that the
image of a bright point $F:=(0,0,-f)$ is a point $(0,0,2d-f)$, $0<f<d$),
remain true if $k_3'\neq k_3$? }

2) {\it Suppose that the slab is not
infinite. Will the diffraction from the boundary of the slab
invalidate the principal conclusion?} 

We will not discuss the second
question in detail, and restrict ourselves to the following
argument. If the size of the slab, which we defined as a maximal
radius $R$ of a cylinder, inscribed into the slab, is much larger
than the wavelength $\lambda_0=\frac{2\pi}{k_0}$,
$k_0=\omega\sqrt{\epsilon_0\mu_0}$, $R\gg \lambda_0$ and $R\gg l$,
where $l$ is the size of the object, then one may neglect the
diffraction of light at the boundary of the slab, provided that the
object is located near $z$-axis, far away from the boundary of the
slab. 

The first question we discuss in more detail. The aim of our
discussion is to conclude that if $\epsilon_1$ or  $\mu_1$
differ slightly from $-1$, so that $n$ is not equal to $-1$ exactly,
then the image of a point $(0,0,-f)$ is no longer a
point. More precisely, not all the rays, passing throught the point
$F:=(0,0,-f)$, pass through the common point of the $z$-axis after
the plane $z=d$. It turns out that the 
point of 
itersection of the ray, passing through $F$ at the angle $\theta_0$ with 
the $z$-axis in the region $z>d$, will depend on $\theta_0$ if 
$|k_3'|\neq |k_3|$. Assume, for example, that
$\mu_1=-1+\delta_\mu$, $\epsilon_1=-1+\delta_\epsilon$,
$\delta_\mu=\delta_\epsilon=\delta$, where
$\delta>0$ is a small number. Then
\be\label{eq11}\begin{split}
k_3'&=\sqrt{\omega^2(-1+\delta_\mu)(-1+\delta_\epsilon)-k_1^2}\approx
\sqrt{\omega^2-k_1^2-\omega^2(\delta_\mu+\delta_\epsilon)}\\
&=k_3\sqrt{1-\frac{2\omega^2
\delta}{k_3^2}}\approx k_3-\frac{\omega^2\delta}{k_3},
\quad k_3=\sqrt{\omega^2-k_1^2}.\\
\end{split} \ee 
Here we have neglected the term of order $O(\delta^2)$,
and assumed that $|k_3|\gg \delta>0$. Thus
$\frac{k_3'}{k_3}=1-\delta_1$, where
$\delta_1:=\frac{\omega^2\delta}{k_3}.$ 

If $k_3=O(\delta)$ then $|k'_3|=O(\delta^{1/2}).$ In this case
$\frac{k_3'}{k_3}$ can take rather arbitrary values depending on the
ratio $\frac{\delta}{|\omega^2-k_1^2|}$, and therefore the deviation
of the corresponding rays from the point $(0,0,2d-f)$ may be much greater 
than $O(\delta)$. 

If $|k_3|\gg \delta$, then
we repeat
the geometrical arguments given for the case $k_3'=k_3$, and get the
following conclusions: the ray, passing through the point
$F:=(0,0,-f)$ and having angle $\theta_0$ with the $z$-axis, hits the
plane $z=0$ at the point $(h_0,0,0)$, $h_0=f\tan\theta_0$,
then it hits the $z$-axis at the point $(0,0,f')$, where $\tan
\theta'=\frac{h_0}{f'}=\frac{k_1}{k_3}$, so $f'=h_0 \cot \theta'$,
and then it hits the plane $z=d$ at the point $(-h_1',0,0)$, where
$\frac{h'_1}{d-f'}=\tan \theta'=\frac{k_1}{k_3'}.$ 
Finally, after
passing the plane $z=d$, this ray hits the $z$-axis at the point
$(0,0,d+z')$, where $z'=h_1'\cot \theta_0$. Therefore,
$$d+z'=d+(d-f\tan\theta_0\cot \theta')\tan\theta'\cot\theta_0=
d(1+\tan\theta'\cot\theta_0)-f.$$
One has $\tan\theta_0=\frac{k_1}{k_3}$,
$\tan\theta'=\frac{k_1}{k_3'}$, so
$\tan\theta'\cot\theta_0=\frac{k_3}{k_3'}=\frac{1}{1-\delta_1}.$\\
\textbf{Conclusion:} {\it The ray, passing through the point 
$F=(0,0,-f)$
under the angle $\theta_0$ with $z$-axis, hits the $z$-axis, after 
passing
the slab, at the point whose coordinate $(0,0, d-f+d\tan\theta'\cot \theta_0)$ 
depends on the angle
$\theta_0$. Therefore, the image of a point source, located at
the point $F$, will not be a point.} 

At a fixed $\omega>0$ the
deviation of the image of a bright point depends on the deviation of
$\frac{k_3}{k_3'}$ from $1$. This implies a possibility to make
changes in $\epsilon_1$, if it is practically easier than to make
changes in $\mu_1$, in order to compensate for the deviations of
$\mu_1$ from $-1$.

\section{On the resolution ability of linear isoplanatic optical 
instruments}
Following L. Mandel'shtam \cite{M}, assume that a linear isoplanatic 
optical
instrument  $\mathcal{S}$ creates an image of an object as follows. A 
bright point
(a point source),  is  located on a plane $P_1$ at a point $y\in 
\R^3$, and
the focal plane is located  to the left of  $\mathcal{S}$.
The bright point produces an image $h(x,y)=h(x-y)$ on the image plane
$P_2$,  
$x\in P_2,$ and $P_2$
is the plane to the right of $\mathcal{S}$. The function
$h=h(x)$ is called the scattering function of the instrument
$\mathcal{S}$. The instrument is called isoplanatic if
$h(x,y)=h(x-y)$. The image of an object $f(x)$, located in the plane
$P_1$ is 
$$u(x)=\int_{P_1}h(x-y)f(y)dy,\quad x\in P_2.$$ 
The problem of
the resolution ability of $\mathcal{S}$ in Rayleigh's formulation is
stated as follows: 

{\it Given two signals, $f_1=\delta(x-a),\
f_2=\delta(x+a)$, where $\delta(x)$ is the delta-function and $a$ is
a vector, $|a|$ is small, will the function
$\frac{h(x+a)+h(x-a)}{2}$
have one or two maxima near $x=0$? If two
maxima can be detected, then, according to Rayleigh, the instrument
$\mathcal{S}$ can discriminate between the two point
sources the distance between which is $2a$.}\\
  This criterion of the resolution ability is very limited: the class
of objects to be discriminated consists of two point sources.
Rayleigh's conclusion is : one cannot discriminate between two point
sources if $a \sim \frac{\lambda}{2}$, because the width of the
function $h(x)$ is of the order of $\lambda$, where $\lambda$ is the
wavelength. 

In [5]-[12] the resolution ability
problem was examined in detail. In \cite{R55} a new definition of
resolution ability of a linear instrument is given. This definition
allows for a very wide class of the input signals, not only for two
point sources (bright points), as in Rayleigh's definition. It takes
into account the noise on the image plane, and it takes into account
the properties of the procedure that reconstructs the input signal
from its noisy image. The principal conclusion in \cite{R55} is :

{\it The resolution ability in Rayleigh's sense can be increased without
a limit by apodization.}

 In other words, the scattering function of a
linear isoplanatic optical instrument can be made by apodization as
close as one wishes to a delta-function on any fixed finite
subdomain of the image plane. This conclusion is justified as
follows. The scattering function is $h(x)=\int_D j(y)e^{-ix\cdot
y}dy,$ where $x\cdot y$ is the dot product of vectors, and $D$ is a
finite region, it is the output pupil of the optical instrument
$\mathcal{S}$, (see \cite{O}, Sec. 5.3), and $j(y)$ is the field
distribution on the output pupil $D$. If one puts a mask on the
output pupil $D$, then the field $j(y)$ is transformed into
$j(y)n(y)$, where $n(y)$ is the refraction coefficient of the
mask. Putting the mask on $D$ results in the change of $h$: the new
$h=h_n:=\int_Dj(y)n(y)e^{-ix\cdot y}dy$. It is proved in
\cite{R42},\cite{R53},\cite{R54}, that, given a bounded domain
$\Delta$ on the image plane, one can find a sequence of 
functions $f_m(y)$ such
that the sequence $\delta_m(x):=\int_Df_m(y)e^{-ix\cdot y}dy$ is a
delta-sequence on $\Delta$,  that is $\lim_{m\to
\infty}\int_{\Delta}\delta_{m}(x-s)\phi(s)ds=\phi(x)$ for any smooth
function $\phi$ vanishing outside $\Delta$. Moreover, the sequence
$f_m(y)$ is given explicitly, analytically, in \cite{R53} and in
\cite{R109}, see also \cite[p.211]{R118}. Therefore one can choose
$n_m(y)$ so that $j(y)n_m(y)=f_m(y)$, and then the
correspionding scattering functions $h_m(x)$ form a delta-sequence
on the finite domain $\Delta$, so that the width of $h_m(x)$ can be made 
as small as one wishes. 

By the Rayleigh criterion, this means that, for
sufficiently large $m$ the resolution ability (in Rayleigh's sense)
can be increased without a limit. In this argument we did not
take into account the noise in the image. It is proved in \cite{R55} 
that it
is possible to increase the resolution ability (in the Rayleigh's
sense) of a linear isoplanatic optical system without a limit by
apodization even in the presence of noisy background in the image
plane. The noise is assumed to be independent identically
distributed with zero mean value and an arbitrary large but fixed
finite variance. Our arguments are valid under the assumption that
the classical Maxwell's equations are valid. We do not discuss
quantum effects in this paper.\\

The author thanks Prof. O.L. Weaver for discussions.

\end{document}